\documentclass[preprint,12pt,authoryear]{elsarticle}

\usepackage{amssymb}
\usepackage{amsmath}

\usepackage{xurl}

\journal{The Electricity Journal}

\begin{document}

\begin{frontmatter}



\title{Ensuring reliability: what is the optimal time for power plant maintenance in Texas as the climate changes?}


\author[inst1]{Hugh Daigle\footnote{Correspondence: daigle@austin.utexas.edu}}
\author[inst2]{Joshua D. Rhodes}
\author[inst3]{Aidan Pyrcz}
\author[inst2]{Michael E. Webber}

\affiliation[inst1]{organization={Center for Subsurface Energy and the Environment, The University of Texas at Austin},
            city={Austin},
            state={Texas},
            country={USA}}

\affiliation[inst2]{organization={J. Mike Walker Department of Mechanical Engineering, The University of Texas at Austin},
            city={Austin},
            state={Texas},
            country={USA}}
            
\affiliation[inst3]{organization={McKetta Department of Chemical Engineering, The University of Texas at Austin},
            city={Austin},
            state={Texas},
            country={USA}}

\begin{abstract}
We analyzed data for the Electric Reliability Council of Texas (ERCOT) to assess shoulder seasons -- that is, the 45 days of lowest total energy use and peak demand in the spring and fall typically used for power plant maintenance -- and whether their occurrence has changed over time. Over the period 1996--2022, the shoulder seasons never started earlier than late March nor later than mid-October, corresponding well with the minimum of total degree days. In the temperature record 1959--2022, the minimum in degree days in the spring moved earlier, from early March to early February, and in the fall moved later, from early to mid-November. Warming temperatures might cause these minima in degree days to merge into a single annual minimum in December or January by the mid-2040s, a time when there is a non-trivial risk of 1-day record energy use and peak demand from winter storms.
\end{abstract}




\end{frontmatter}

\section{Introduction}
\label{sec:introduction}


It is widely recognized that climate change will adversely affect reliable electricity delivery worldwide due to more frequent extreme weather events, increased demand for space cooling, and the relative vulnerability of power generation infrastructure to increasing temperatures \citep{vanvlietetal2012,zamudaetal2018,coffelmankin2021}. Unreliable electricity delivery results in loss of life during hot and cold weather, and has economic consequences: for example, blackouts during Winter Storm Uri in the United States in February 2021 caused more than 200 detahs and about \$155 billion in economic losses \citep{busbyetal2021}. A large number of recent studies have looked at the ways in which climate change will affect peak electricity demand in different parts of the world, focusing on both summer and winter peaks \citep{parkpoometal2008,chenlie2010,ahmedetal2012,sathayeetal2013,bartoschester2015,trotteretal2016,auffhammeretal2017,burilloetal2019,fanetal2019,chabounietal2020,garridoperezetal2021,shafferetal2022,romittiwing2022}. However, variations in the intensity of the hottest and coldest parts of the year are not the only consideration for predicting future electricity reliability. As average global temperature increases, spring is starting earlier and summer is lasting longer as indicated by several different metrics including temperatures, plant behavior, bird migration and nesting, and the Asian monsoon \citep{schwartzetal2006,thompsonclark2008,penaortizetal2015,lehikoinenetal2019,shipleyetal2020,hoetal2021}. The effect of these climate change-induced variations in seasonality on electricity reliability have been addressed mainly in terms of meeting increased electricity demand over a longer period during the hot time of the year \citep{hamletetal2010,vanruijvenetal2019,hilletal2021}. However, a phenomenon that has not been explored to date is how periods of minimum electricity demand -- which are optimal times for scheduled maintenance of power plants -- have changed and will continue to change in the future. Here, we use data from the U.S. state of Texas to show how earlier onset of spring and later onset of fall affect the timing of the periods with lowest electricity demand, raising the risk of shorter maintenance windows that could reduce power plant reliability. While our results are specific to Texas, the conclusions are broadly applicable as the lengthening of summer due to climate change is a globally observed phenomenon.

Texas is the largest producer and consumer of electricity in the United States. In 2020, generation and consumption were nearly twice as high as the next-highest states \citep{eia_sep}. The high demand is due to the state's large industrial base, which in 2020 accounted for 53.9\% of total electricity consumption \citep{eia_texas_energy_profile}; hot climate and weak building efficiency standards, which lead to significant space cooling; and robust population growth, with a 15.9\% increase from 2010 to 2020 \citep{census} and a further 70\% increase forecast by 2070 \citep{tx_state_water_plan}. The Electric Reliability Council of Texas (ERCOT), which is the grid that serves about 90\% of Texas electricity, projects that peak demand during the summer could increase from 81.6 GW in 2024 to 88.7 GW in 2033, or by about 9\% \citep{ercot_cdr}. The summer of 2023 saw large peak demand, includingan all-time record of 85.5 GW on August 10.

Power plants are complex machines with many moving parts that require going off-line for maintenance, similar to how a light duty internal combustion vehicle must turn off for periodic oil changes. However, there are limited periods of time that are conducive for power plants to take those extended outages without compromising grid reliability. Predicting the optimum times for this preventative maintenance requires consideration of many different factors, including generation ramp rate, availability of personnel and resources, transmission line capacity, demand, and reliability \citep{frogeretal2016}. Because the ERCOT grid is sensitive to temperature-driven spikes in power use and peak demand from heating and cooling needs \citep{rhodesetal2011,alipouretal2019,shafferetal2022}, ERCOT allows electricity generators to schedule planned maintenance as long as the total amount of generation capacity taken offline during a particular period does not exceed the forecasted maximum daily resource planned outage capacity (MDRPOC). MDRPOC data are forecasted for 7 days into the future. Projections farther into the future are based on historical data, specifically the 50$^{th}$ percentile of past load profile data with MDRPOCs from the past 3 years for the summer and winter months \citep{ercot_mdrpoc}. This approach has the effect of allowing most planned power plant maintenance in the spring or fall -- the traditional shoulder seasons -- when temperatures are historically mild and electricity demand is generally lower.

As a result of climate change, the onset of springtime biological activity has moved earlier in the year by 9.3 days in the Central Plains and by 18.8 days in the southwestern United States since 1950 \citep{crimmins2019}, and the length of the frost-free season in Texas has increased by 12--24 days over the same period \citep{zhang_frost_free_days,modalaetal2017}. It follows that the shoulder seasons have likely changed their duration and timing over the same period, and will change in the future as the planet continues to warm. ERCOT does not explicitly consider climate change in their planning, which could lead to reliability concerns in the future \citep{leedessler2022}. Maintaining stable, reliable grid conditions requires understanding how changing season onsets have been reflected in temperature data and electricity demand, and how those might continue to change in the future. We analyzed temperature and electrical grid load data for the ERCOT service area to assess how shoulder seasons have varied and provide guidance on how they might continue to vary in the future.

\section{Background}
Over the period 1895--2021, the average daily maximum and minimum temperatures in Texas increased by 0.8 Celsius degrees, and overall average temperature is projected to increase by another 1 Celsius degree by 2036 relative to the 1999-2020 average. While the absolute number of days above 100$^\circ$F (38$^\circ$C) varies geographically across the state, this projected temperature increase means that locations across Texas would experience 40\% more days above 100$^\circ$F (38$^\circ$C) \citep{climatereport2021}. Further, the timing of the onset of spring and autumn have been changing and are expected to continue to do so in the future \citep{crimmins2019,zhang_frost_free_days,modalaetal2017}. This shift is consistent with other work demonstrating earlier snow disappearance and leaf growth in the spring and later snow appearance and leaf loss in the fall at more northern latitudes \citep{creedetal2015,groganetal2020}. These trends are projected to continue in the future as average temperatures increase.

Electricity demand in the ERCOT service area has changed over the past few decades due to the combined effects of climate change, population growth, efficiency increases in domestic cooling, and increases in the proportion of households using electric heating \citep{whiteetal2021,leedessler2022,skilesetal2023}. Both \citet{leedessler2022} and \citet{skilesetal2023} demonstrated that climate change has specifically increased summer peak demand, and that this trend is expected to continue in the future because residential cooling drives spikes in electricity demand on hot summer days in Texas \citep{whiterhodes2019}. It follows, then, that residential electricity demand should also respond to earlier onset of spring weather and later onset of fall weather, which in turn should drive changes in the timing of the shoulder seasons.

In 2022, the shoulder seasons were predicted by ERCOT to occur from early March to late May, and from mid-September to late November \citep{rickerson_shoulder}. This follows the reported planned outages from 2021, with no planned outages between June 10 and September 15 and the winter having on average 18\% the outages of the spring shoulder season (3,000 MW versus 17,000 MW). \citet{osheaetal2021} analyzed all generation resource outages (planned and forced) reported by ERCOT for the period 2015--2020 and similarly showed that outages were about 15 GW higher during March--May and October--November relative to the winter and summer averages, respectively, thus demonstrating how planned outages drive overall outages in shoulder seasons. However, it is not clear that these time periods are the optimal shoulder seasons in terms of reliably meeting demand. Anomalously hot or cold weather can occur during these predicted shoulder seasons. When projected electricity demand is expected to approach generation capacity, ERCOT issues a conservation request in an effort to reduce demand. Between January 2008 and July 2022, ERCOT issued 44 area-wide conservation requests \citep{ercot_conservation}. Most were during the summer or winter, but 4 were issued between March 1 and May 31 and 1 between September 15 and November 30, which means that 11\% of these conservation requests were issued during the periods ERCOT defines as shoulder seasons. In April 2006, generation was so constrained that ERCOT instituted rolling blackouts across its entire service area as 20\% of generation was offline for maintenance \citep{aptetal2006}. Notably, since the 1980s, area-wide rolling blackouts have only occurred 4 times, with 3 during the winter and 1 during the spring shoulder season \citep{munce_tex_trib}. Here, we demonstrate quantitatively how climate change has affected the shoulder seasons since the mid-20\textsuperscript{th} century, and predict how these changes may continue over the 21\textsuperscript{st} century.

\section{Data and methods}
\label{sec:methods}
In this analysis, we sought to assess how the timing of shoulder seasons is changing and could continue to do so in the future with climate change. We used three metrics to define shoulder season: daily electricity use, daily peak demand, and heating- and cooling-degree days. Shoulder seasons in the spring and fall were defined as the 45-day periods with the lowest average of each of these three metrics. Our choice of 45 days as the length of the optimal window was motivated by a rule revision, Nodal Protocol Revision Request 1108 (NPRR 1108), that ERCOT proposed in 2021. The ERCOT Technical Advisory Committee recommended setting minimum MRDPOC levels during the shoulder seasons, with maximum MRDPOC March 15-May 1 and October 15-November 30 \citep{jointcommentersI,samsTACadvocate}. This recommendation had the support of electricity generators and other stakeholders \citep{richmondTCPAresponse}. While this part of NPRR 1108 was ultimately not included in the final, revised rule \citep{ercotboardreport}, it does give an indication of the practical length of the ideal periods for power plant maintenance.

\begin{figure}
    \centering
    \includegraphics{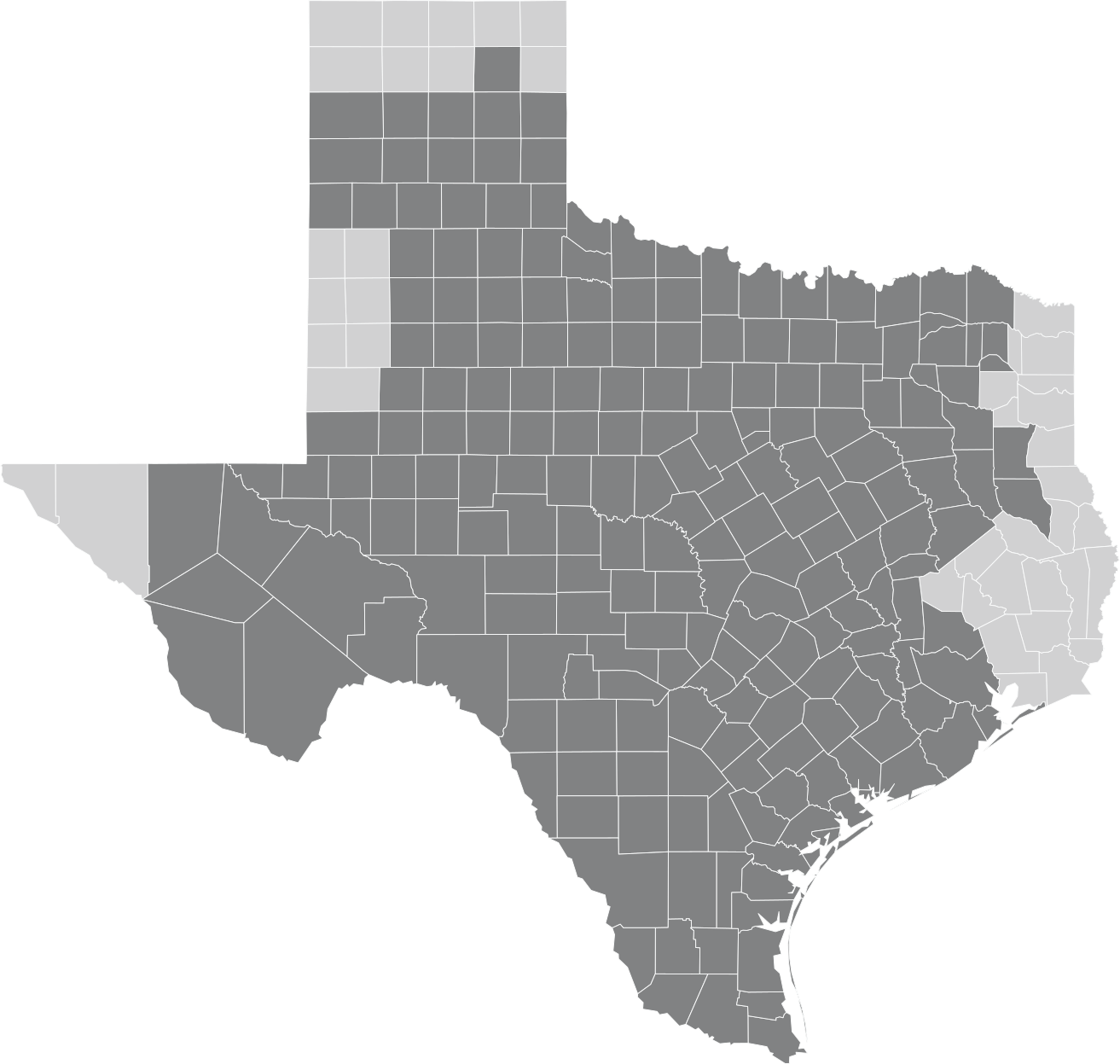}
    \caption{Map of Texas with ERCOT service area in dark gray. Base map from \protect\url{https://freevectormaps.com}}
    \label{fig1}
\end{figure}

Our analysis considered the ERCOT service area, which encompasses most, but not all, of Texas (Fig.~\ref{fig1}). Electricity use data came from ERCOT, which has hourly load data publicly available for every day from 1 January 1996 to the present with the exception of 2001. For temperature data, we used the European Centre for Medium-Range Weather Forecasts Reanalysis 5 (ECMWF ERA5) of atmospheric temperature dating back to 1959 \citep{era5}. ERA5 provides hourly temperature data at a spatial resolution of 0.25$^\circ$ at various atmospheric levels; we used the temperature at 2 meters above ground level. 

We used total degree days to define temperature-based shoulder seasons. Degree days represent the deviation of the average temperature for a particular day above or below a baseline and are used as an indication of energy demand for heating or cooling. In calculating degree days, population-weighted temperature is typically used because energy demand will be concentrated in areas with higher population density, and therefore the temperatures in those areas will more strongly influence load on the electric grid \citep{quaylediaz1980,taylor1981}. We determined population-weighted temperatures following \citet{leedessler2022}. Temperature data were weighted by population using data from the Center for International Earth Science Information Network (CIESIN), which is gridded at the same resolution as the temperature data \citep{ciesin}. The CIESIN data have updates every 5 years from 2000 to 2020, which we used for this date range. For dates prior to 2000, the 2000 population weights were used.

\begin{figure}
    \centering
    \includegraphics{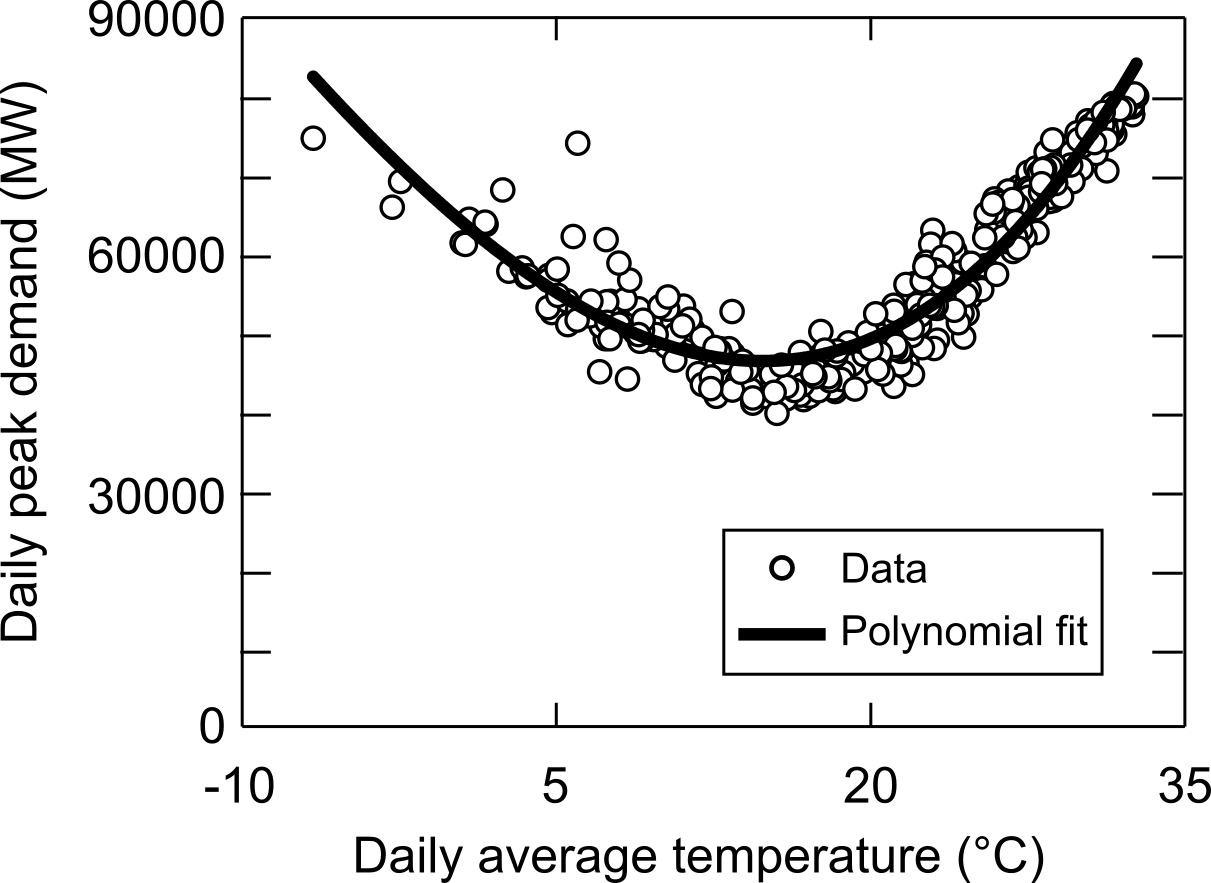}
    \caption{Illustration of determining $T_0$. Peak demand is plotted as a function of daily average temperature, and a $3^{rd}$ order polynomial is fit to the data. $T_0$ is the temperature at the minimum peak demand predicted by the polynomial. Data shown are from 2022.}
    \label{fig2}
\end{figure}

To process the power load data, we computed the daily total use (energy, kWh) and peak demand (power, kW), and then determined the 45-day period with the lowest average daily total use and peak demands in the first half (January--June) and second half (July--December) of the year. To process the temperature data, we first computed the average daily temperature $T_{avg}$ across the entire ERCOT service area using population-weighted temperatures. We then determined total degree days $DD$, which we define as

\begin{equation}
    DD = \begin{cases}
        T_{avg} - T_0, & T_{avg} \geq T_0\\
        T_0 - T_{avg}, & T_{avg} < T_0
        \end{cases},
        \label{eq1}
\end{equation}

\noindent where $T_0$ is a reference temperature. $T_0$ was determined by plotting daily peak demand versus daily average temperature, fitting a $3^{rd}$ degree polynomial to the data, and finding the average temperature corresponding to the minimum daily peak demand predicted from the polynomial fit (Fig.~\ref{fig2}). Polynomial coefficients and $T_0$ values are listed in Table A3 in the Supporting Information. We made one plot for each year where data were available, and $T_0$ was taken as the average over all available years. The dates we report are the onsets of the 45-day periods of minimum total use, peak demand, and degree days.

To assess whether changes over time in the onset of shoulder seasons were statistically significant, we conducted a least-squares linear regression and found the rate of change along with its standard error. We then computed the probability that the rate of change indicated a shift earlier in the spring (rate of change $<0$) or later in the fall (rate of change $>0$).

\section{Results and Discussion}
\subsection{Trends in temperature-defined shoulder seasons}
Since the temperature record we used extends farther back than the electricity usage dataset, it provides a useful proxy for long-term trends in energy demand in the ERCOT service area. From 1959 to 2022, the spring shoulder season moved earlier in the year by 2.4 days per decade (Fig.~3a), while the fall shoulder season moved later by 1.1 days per decade (Fig.~3b). The rate of change of the spring shoulder season is consistent with the change in onset of springtime biological activity in the southwestern United States reported by \citet{crimmins2019} (2.7 days per decade). Based on the standard error of the slopes of our regression lines, these shifts occurred with probabilities of 99\% and 91\% in the spring and fall, respectively. During the 64 years in this dataset, the spring shoulder season never started later than March 12 and the fall shoulder season never started earlier than September 29.

\begin{figure}
    \centering
    \includegraphics{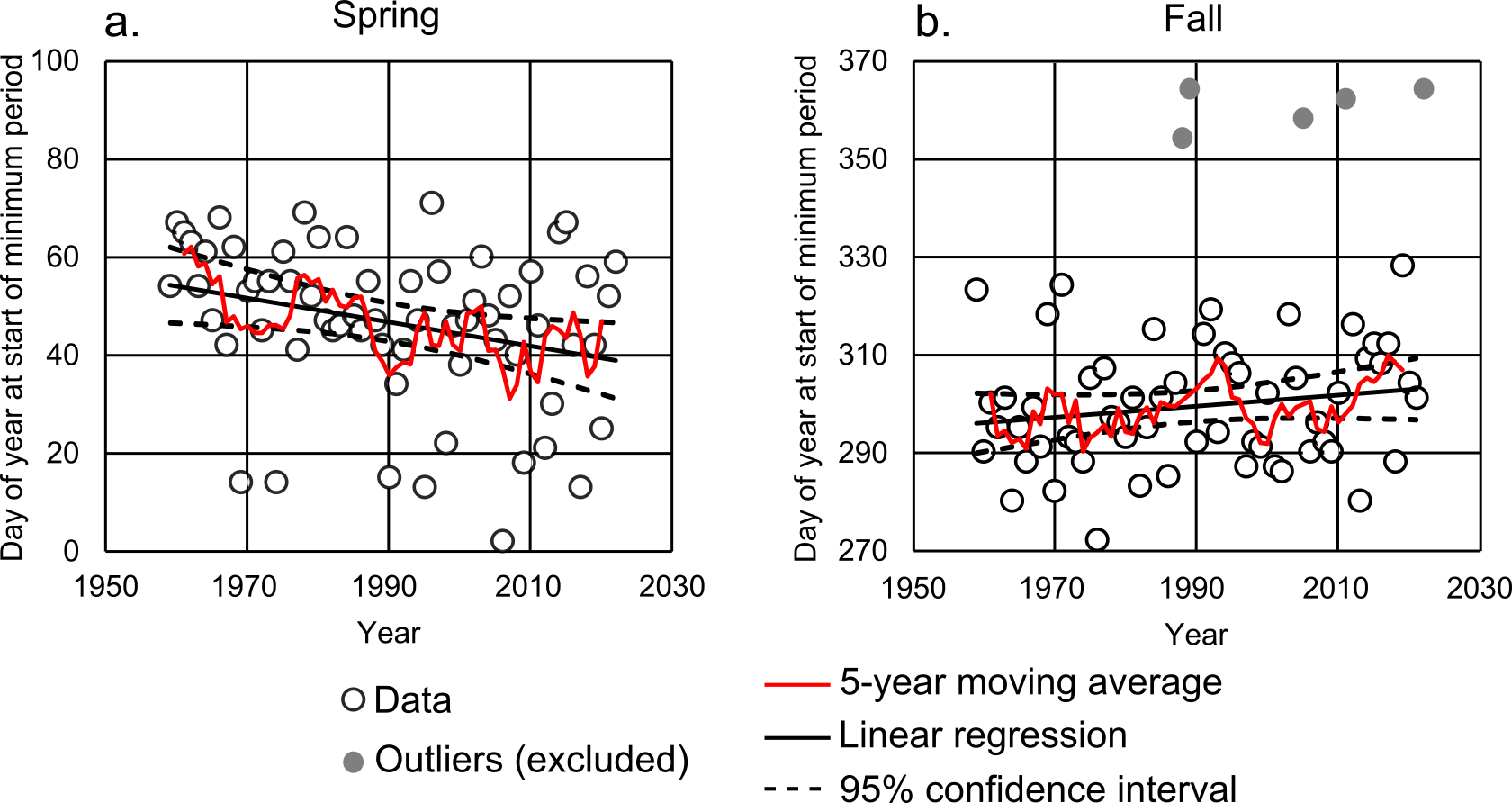}
    \caption{Start of spring (a) and fall (b) shoulder seasons defined by degree days. 5-year moving averages are included (red lines), as well as linear regression lines with 95\% confidence intervals. Note that 5 outliers were excluded from the regression in the fall.}
    \label{fig3}
\end{figure}

\subsection{Trends in electricity-defined shoulder seasons}

The 45-day periods of minimum total energy use (kWh) and minimum peak demand (kW) in the spring have shifted slightly earlier in the year since 1996 (Figs.~4a,b), by 2.0 days per decade for total energy use and 1.0 days per decade for peak demand. Due to the shorter time series compared to the temperature data, the probability that these shoulder seasons shifted earlier is lower than observed in the temperature data, at 69\% for total energy use and 60\% for peak demand. These spring shoulder seasons never began later than March 14.

In the fall (Figs.~4c,d), least squares regression indicates that the shoulder season defined by total energy use moved later in the year by 0.057 days per decade while that defined by peak demand moved later by 0.027 days per decade. However, these changes are very small and uncertain. The probability that a shift later actually occurred is 51\% for total energy use and 50\% for peak demand, thus giving nearly equal probabilities that the fall shoulder seasons shifter earlier or later. The fall shoulder seasons never started earlier than October 6 for total energy use and October 12 for peak demand.

To summarize these results, the 45-day periods of minimum total energy use and peak demand started in February--early March and early October--early November over this time period. These results are consistent with the proposed period of maximum allowable MRDPOC proposed by the ERCOT Technical Advisory Committee \citep{samsTACadvocate}.

\begin{figure}
    \centering
    \includegraphics{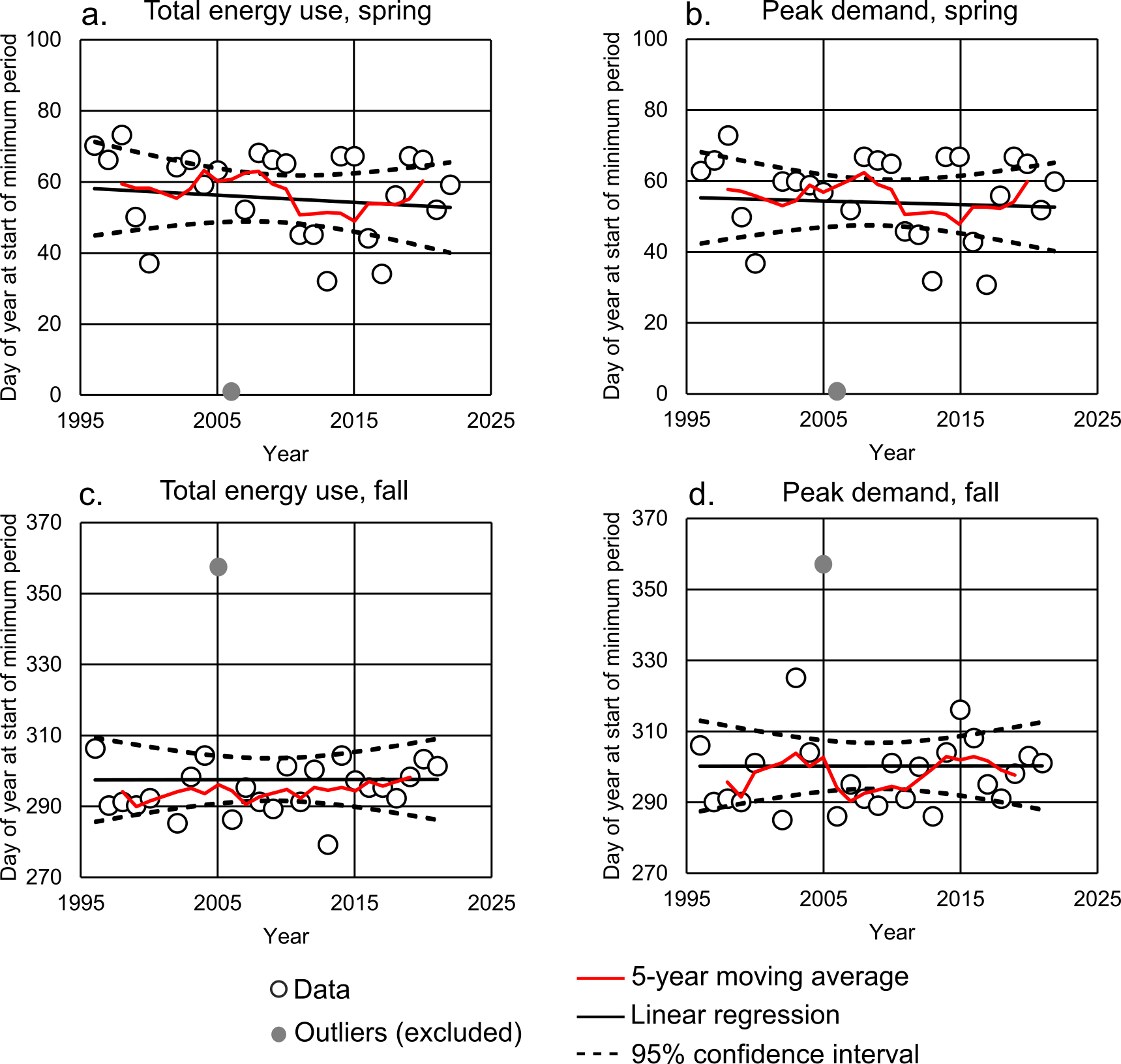}
    \caption{Shoulder seasons defined by electricity. (a) Spring shoulder season defined by total energy use. (b) Spring shoulder season defined by peak demand. (c) Fall shoulder season defined by total energy use. (d) Fall shoulder season defined by peak demand. 5-year moving averages are included (red lines), as well as linear regression lines with 95\% confidence intervals.}
    \label{fig4}
\end{figure}

\subsection{Correlation between degree days and electricity demand}

The existence of a correlation between minimum degree days and minimum electricity demand implies that climatological data could and should be used to predict changes in shoulder season in the future. ERCOT does not incorporate long-term climate forecasts into their demand forecasts \citep{leedessler2022}, even as many studies have concluded that climate change will affect electricity demand in the future \citep{fanetal2019,auffhammeretal2017,emodietal2018,garridoperezetal2021}.

In the spring, the shoulder season defined by degree days has a good correlation with those defined by total energy use and peak demand, but only when temperature-defined shoulder seasons starting earlier than February 14 are excluded (Pearson correlation coefficients of 0.80 and 0.76, respectively) (Figs.~5a,b). Similarly, in the fall, there is good correlation when temperature-defined shoulder seasons starting later than November 25 are excluded (Pearson correlation coefficients of 0.71 for total energy use and 0.74 for peak demand) (Figs.~5c,d). Thus, we conclude there is a weaker correlation between statewide average temperature and electricity usage during the winter months (December, January, and February), which could be driven by greater spatial variability in temperatures across the ERCOT service area in winter. Importantly, as spring weather starts earlier in the year and fall weather persists later in the year, there will be less of a relationship between electricity demand and climate, which will complicate planning efforts for grid reliability.

\begin{figure}
    \centering
    \includegraphics{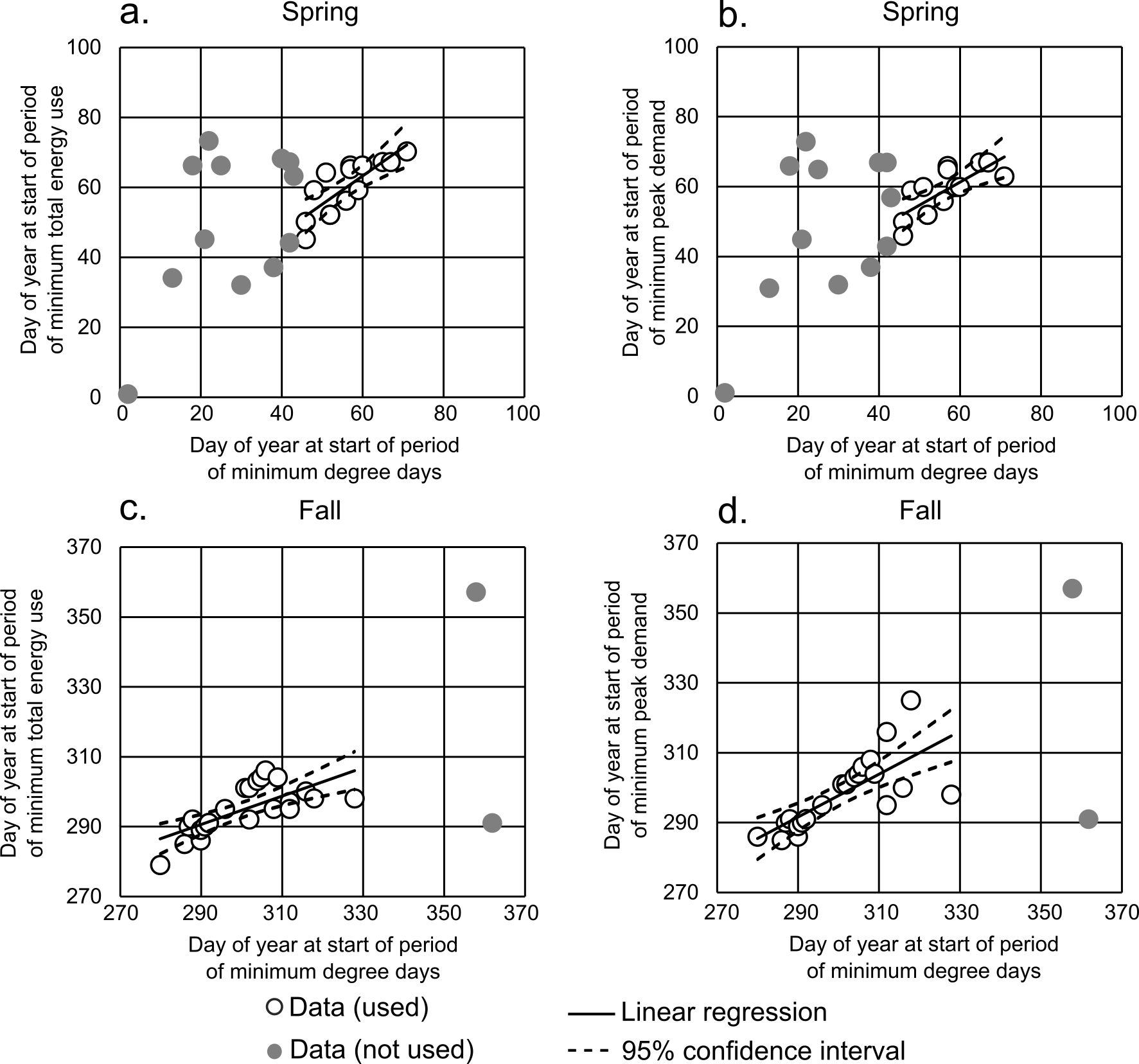}
    \caption{Correlations between start of 45-day periods of minimum degree days, minimum total energy use, and minimum peak demand during spring (a, b) and fall (c, d). The gray data points occur before or after the dates mentioned in the text.}
    \label{fig5}
\end{figure}

\subsection{Effect of renewable energy generation}

Renewable energy generation has increased considerably in ERCOT over the last two decades. From 2007 to 2022, wind, solar, hydroelectric, biomass, and nuclear generation together grew from 17\% of energy generated to 41\%, with wind alone accounting for 25\% of generation in 2022. Wind generation in Texas tends to be the greatest in April and May due to higher wind speeds during those months \citep{eia_texas_wind}. Because our analysis focuses on optimal timing for power plant maintenance, we assessed whether electricity demand in ERCOT followed the same trends with temperature and time when the energy use and peak demand being met by non-thermal generation (wind, solar, hydroelectricity) was removed. Fuel mix data from ERCOT are only available from 2007 onward, so the time series we used was shorter than the overall dataset.

We determined that the shoulder seasons determined by total energy use have a 15\% probability of having shifted earlier in the spring and an 86\% probability of having shifted later in the fall, while those determined by peak demand have a 12\% probability of having shifted earlier in the spring and a 97\% probability of having shifted later in the fall (Fig.~A10). However, we caution that the short time series we used results in trend line slopes in spring shoulder season that are not statistically different from those we obtained by including non-thermal generation, even at a 68\% confidence interval (1 standard deviation). Indeed, the removal of non-thermal generation moved the onset of spring shoulder season an average of 1.5 days later based on total energy use and 0.6 days earlier based on peak demand. In the fall, on the other hand, shoulder seasons started on average 10.8 days later and 8.3 days later.

When compared against the shoulder seasons defined by minimum degree days, we find that the electricity data with non-thermal generation removed follow a very similar trend as before, with electricity-defined shoulder seasons showing good correlation with temperature-defined shoulder seasons when the temperature-defined shoulder season starts later than February 14 in the spring and earlier than November 25 in the fall (Fig.~A11). The Pearson correlation coefficients in the spring are 0.85 and 0.81 for total energy use and peak load, and in the fall are 0.66 and 0.61.

On the basis of this analysis, we conclude that the seasonal variability in non-thermal generation does not have a significant effect on the spring shoulder season, and may cause the fall shoulder season to shift even later in the year. We observe that the changes in temperature-defined shoulder seasons still track the electricity-defined shoulder seasons except during the winter months.

\subsection{Future outlook}

Over the time series we analyzed for degree days (1959--2022), we found a high likelihood that the onset of the 45-day period of minimum degree days shifted earlier in the spring and later in the fall by 2.4 days/decade and 1.1 days/decade, respectively. However, the likelihood of such shifts in total power use or peak demand was much lower, which is likely due to the shorter time series of electricity data (1996--2022 with 2001 omitted). How do we expect these shoulder seasons to move in the future as a result of climate change?

The annual average temperature across the ERCOT service area has increased since 1959 (Fig.~6a) and is expected to continue to do so \citep{climatereport2021}. To predict future temperatures, we used the Community Earth System Model Large Ensemble Community Project 2 (LENS2) \citep{lens2}. LENS2 is an ensemble of 100 different simulations that start at different years in the 9\textsuperscript{th} and 10\textsuperscript{th} centuries with different perturbations in atmospheric potential temperature, Atlantic Meridional Overturning Circulation state, and sea surface height in the Labrador Sea. Future climate projections use the SSP3-7.0 forcing scenario, which is a high-emissions scenario \citep{IPCC_2022}.

LENS2 outputs 2 m land temperatures at several temporal resolutions. We used average monthly temperatures to determine the annual average temperature over all ensemble members in the ERCOT service region. The annual average LENS2 temperatures displayed a systematic high bias relative to the ERA5 temperatures, so we applied a correction by minimizing the misfit in a least-squares sense between the ERA5 temperature and LENS2 annual ensemble means for the period 1959--2022:

\begin{equation}
    T_{\text{L,corr}} = 0.92T_{\text{L}} + 1.0,
        \label{eq2}
\end{equation}

\noindent where $T_{\text{L}}$ is the raw LENS2 ensemble mean and $T_{\text{L,corr}}$ is the corrected LENS2 ensemble mean. Fig.~6a shows the corrected LENS2 temperatures $\pm2$ standard deviations. Notice how there is less year-to-year variability in the LENS2 temperatures, likely due to the lower spatial resolution (1$^\circ$ versus 0.25$^\circ$ for ERA5).

\begin{figure}
    \centering
    \includegraphics{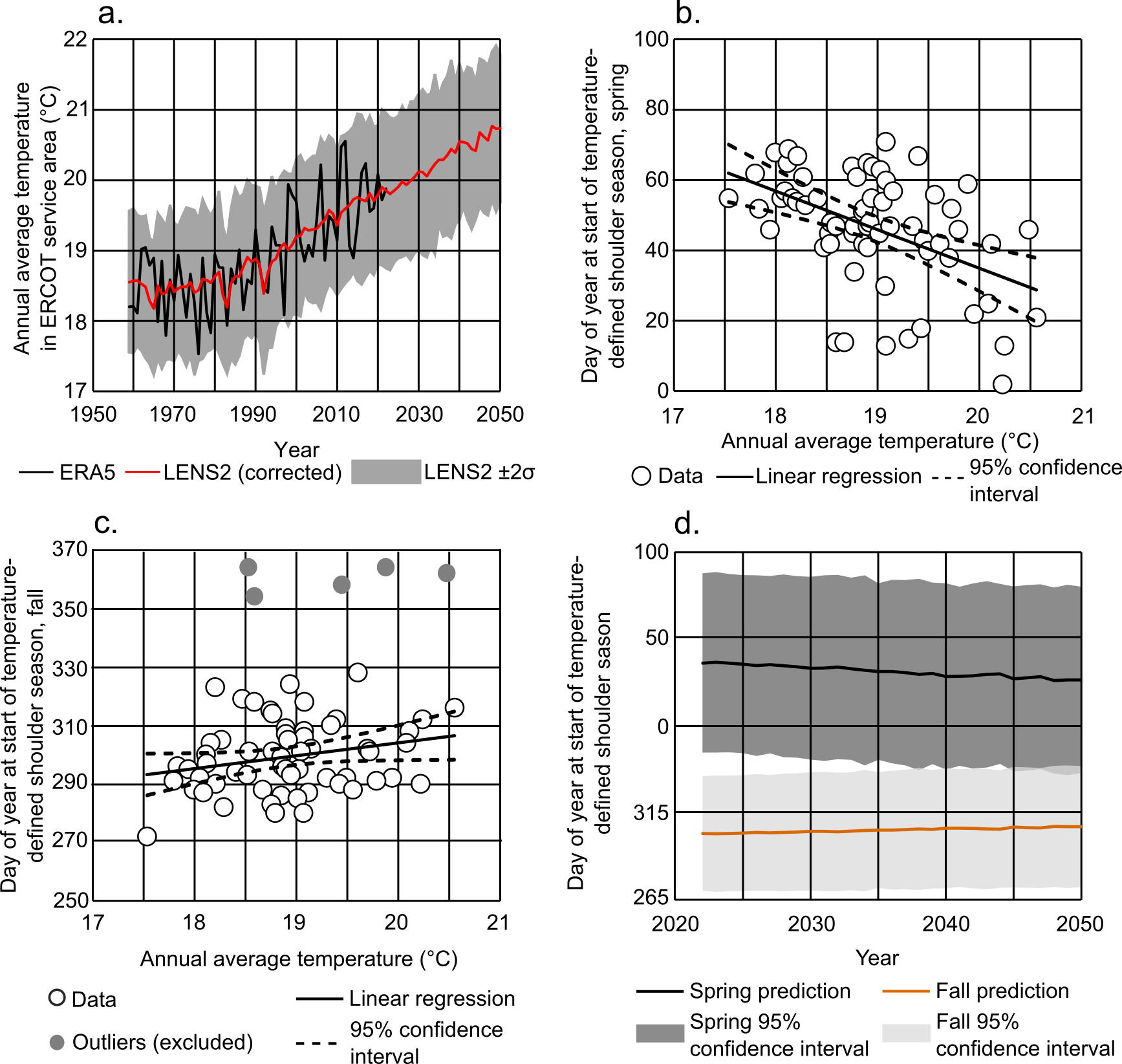}
    \caption{(a) Average annual temperature over ERCOT service area from ERA5 (not population-weighted) and corrected LENS2 temperature $\pm2$ standard deviations, which shows a generally increasing trend. (b) Correlation between ERA5 average annual temperature and day of onset of temperature-defined spring shoulder season, which shows an earlier start date over time. (c) Correlation between ERA5 average annual temperature and day of onset of temperature-defined fall shoulder season, which shows a later start date over time. (d) Predicted onset of temperature-defined shoulder seasons using corrected LENS2 predictions.}
    \label{fig6}
\end{figure}

We observe correlations between average annual temperature and the onset of shoulder seasons. As the average annual temperature (not population-weighted) has increased, the temperature-defined spring shoulder season has moved earlier (Fig.~6b), while the temperature-defined fall shoulder season has moved later (Fig.~6c). Using the trends shown in Figs.~6b and 6c and the LENS2 temperature predictions, we show that the temperature-defined spring and fall shoulder seasons may overlap consistently (defined as overlap of the 95\% confidence intervals) in the mid-2040s as a result of this warming (Fig.~6d). In this scenario, there might be only a single shoulder season centered in December and January.

Such a shift will make shoulder season for electricity demand more unpredictable. As shown in Fig.~5, there is little correlation between the onset of the temperature-defined spring shoulder season and those defined by total electricity demand and peak load when the temperature-defined shoulder season starts earlier than February 14. This lack of correlation is due to the spatial variability of temperature being much higher during this time of year compared with the rest of the year (Fig.~7). Scheduling power plant maintenance in January and February is very risky because extreme cold can occur during these months -- for example Winter Storm Uri in 2021 -- and a grid with a significant amount of generation offline for maintenance would struggle to keep up with demand. At the same time, restricting maintenance to March through May means that maintenance periods will no longer coincide with periods of low expected demand. One solution would be to restrict maintenance to the fall shoulder season, which we show should occur in October and November, but it may be unreasonable to expect power plants to be able to forgo spring maintenance.

\begin{figure}
    \centering
    \includegraphics{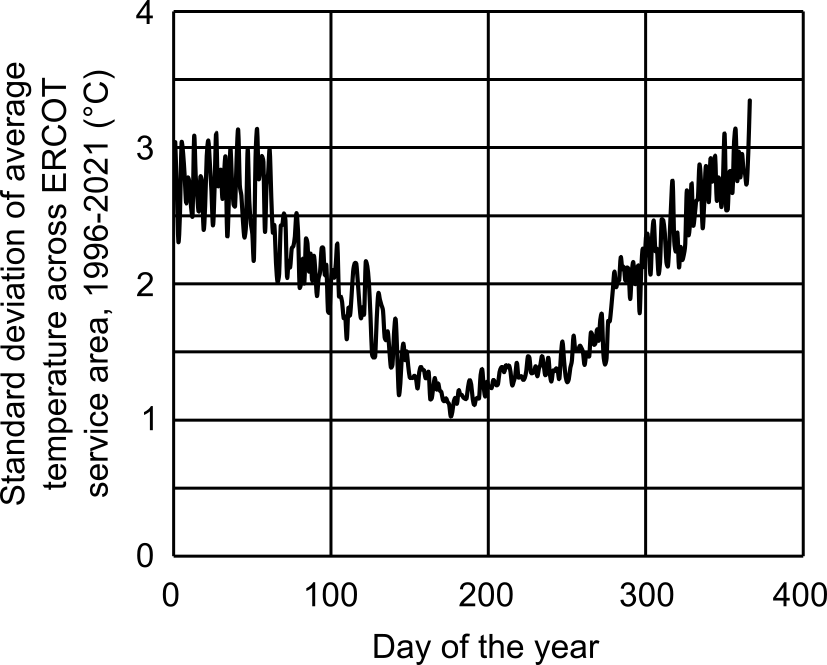}
    \caption{Standard deviation of daily average temperature (not population-weighted) across the ERCOT service area averaged over the interval 1996--2021.}
    \label{fig7}
\end{figure}

Electricity demand during shoulder seasons is also a concern as it dictates how much generating capacity must be kept online to ensure grid reliability. Our work shows that the timing of shoulder seasons is related to minima in heating- and cooling-degree days, and \citet{leedessler2022} similarly demonstrated how temperature affects electricity demand. However, electricity demand in ERCOT is also influenced by other factors, including population growth, electrification of residential heating \citep{whiteetal2021,skilesetal2023}, electrification of oilfield operations \citep{linetal2022}, and even cryptocurrency mining \citep{lee2023quantification}. Over the period 1959--2022, the average degree days during the temperature-defined spring and fall shoulder seasons have changed very little (Figs.~8a,b). On the other hand, total energy use and peak demand during their respective shoulder seasons have generally increased since 1996 (Figs.~8c,d). These trends must be taken into account for reserve forecasting during shoulder seasons.

\begin{figure}
    \centering
    \includegraphics{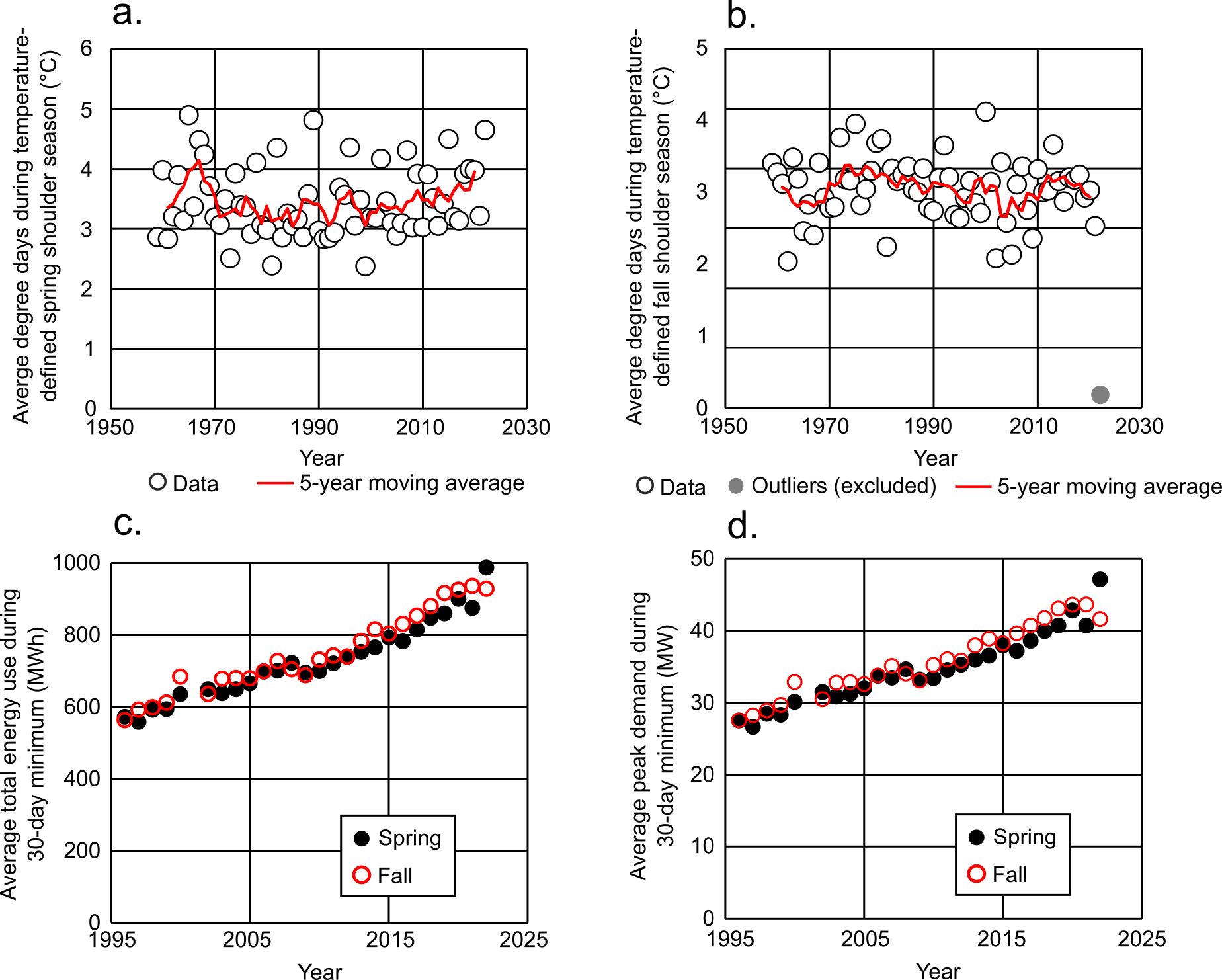}
    \caption{(a) Average degree days during temperature-defined spring shoulder season by year. (b) Average degree days during temperature-defined fall shoulder season by year. (c) Average total energy use during spring and fall shoulder seasons by year. (d) Average peak demand during spring and fall shoulder seasons by year.}
    \label{fig8}
\end{figure}

We assessed the probability that demand could exceed supply during December and January if these become the ideal time of year for power plant maintenance. ERCOT has published generation outage data at 15-minute intervals since May 2021. We combined these data with reported generation output for 2020--2022. Fig.~9 shows histograms of generation output for the ERCOT-defined spring and fall shoulder seasons (March 15--May 1 and October 15--November 30), the 45-day periods of minimum peak demand in spring and fall for each year, and January and December. Peak demand for each period is also indicated. During the ERCOT-defined shoulder seasons, generation is matched very closely with demand, particularly in the spring, which is consistent with ERCOT's strategy of balancing approval of planned outages with forecast demand \citep{rickerson_shoulder}. In the 45-day periods of peak demand, on the other hand, there is excess generation, and more planned outages could be scheduled for these intervals. January and December show a very close balance between supply and demand, and increasing planned outages during these months would present a challenge.

\begin{figure}
    \centering
    \includegraphics{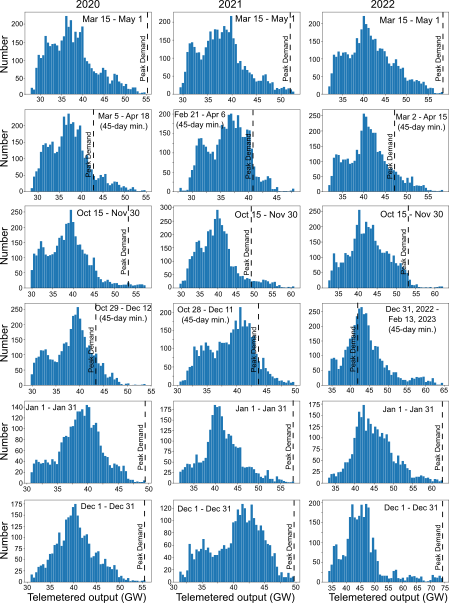}
    \caption{Histograms of generation output for ERCOT in 2020--2022 over different time periods. The peak demand for each time period is indicated with a vertical dashed line. When telemetered output exceeds peak demand, additional outages could be scheduled. When peak demand coincides with the maximum telemetered output, supply and demand are balanced and additional planned outages would risk demand exceeding supply.}
    \label{fig9}
\end{figure}

Table 1 lists the average generation outages in 2022 for the ERCOT-defined spring and fall shoulder seasons, the 45-day periods of minimum peak demand in the spring and fall, and January and December. These numbers include both forced and scheduled outages, but we assume that the larger values during the shoulder seasons reflect a greater amount of scheduled outages. The average outages for December and January together were 16.6 GW while the average for March 15--May 1 and October 15--November 30 together were 22.1 GW. If we assume that the difference between these values (5.5 GW) represents the incremental outages associated with maintenance during the shoulder seasons, we can determine the extent of a deficit in generation that would occur during a shoulder season centered on December to January. Table 2 shows the percentage of ERCOT-wide demand in January and December of 2020--2022 that would have exceeded the maximum generation capacity for those months if an additional 5.5 GW were offline for planned outages. We show that as low as 1.75\% or as high as 23.1\% of demand would not be met in this scenario. If planned maintenance is unable to be performed at other times of year, shifting planned outages to December and January will increase the risk of outages unless additional system-wide generation capacity is added.

\begin{table}
\centering
\begin{tabular}{ |c|c| } 
 \hline
 Time period & Average outages (GW) \\ 
 \hline
  January 1--January 31 & 10.3 \\
 \hline
 March 15--May 1 & 24.0 \\ 
 \hline
 October 15--November 30 & 20.2 \\ 
 \hline
 March 2--April 15 & 22.8 \\
 \hline
 December 1--December 31 & 12.6 \\
 \hline
 December 31, 2022--February 13, 2023 & 8.64 \\
 \hline
\end{tabular}
\caption{Average generation outages for different periods in 2022 and 2023.}
\label{table:1}
\end{table}

\begin{table}
    \centering
    \begin{tabular}{ |c|c| }
        \hline
        Month & \% of demand not met \\
        \hline
        January 2020 & 10.2 \\
        \hline
        December 2020 & 1.75 \\
        \hline
        January 2021 & 2.02 \\
        \hline
        December 2021 & 23.1 \\
        \hline
        January 2022 & 2.02 \\
        \hline
        December 2022 & 2.42 \\
        \hline
    \end{tabular}
    \caption{Percent of demand that would not have been met with an additional 5.5 MW of planned outages for maintenance during December and January.}
    \label{table:2}
\end{table}

\section{Conclusions}

Understanding when minima in electricity demand occur throughout the year, and how the occurrence of those minima will be affected by climate change, is essential for scheduling planned generation facility maintenance and preserving the overall resilience of the electrical grid. In the ERCOT grid, maintenance is typically scheduled in March to May and September to November. However, these time periods may not be optimal, as evidenced by recent events including anomalously hot weather and high electricity demand during the spring shoulder season. We defined shoulder seasons as the 45-day period with lowest average degree days, average energy use, or average peak demand. Over the period 1959--2022, the temperature-defined shoulder seasons shifted earlier by 2.4 days per decade in the spring and later by 1.1 days per decade in the fall. The spring shoulder season never started later than March 12 and the fall shoulder season never started earler than September 29.

Over the shorter period 1996--2022, the shoulder seasons for total energy use never started later than March 14 or earlier than October 6; while those for peak load never started later than March 14 or earlier than October 12. We observed a slight shift earlier in the spring shoulder seasons but not in the fall shoulder seasons. The period of minimum degree days was well correlated with the onset of shoulder seasons defined by total energy use and peak demand when those shoulder seasons began later than February 14 in the spring and any time in the fall, but very poorly correlated for earlier shoulder season onsets in the spring. This lack of correlation is likely attributable to increased variability of temperature over the ERCOT service area during January and February.

The onset of temperature-defined shoulder seasons is correlated with annual average temperature across the ERCOT service area in both the spring and fall. Using corrected LENS2 temperature predictions, these shoulder seasons could merge into a single shoulder season in December and January by the mid-2040s. A single merged season potentially poses problems for balancing electricity generation maintenance with grid stability as shoulder seasons become much more difficult to predict when minimum degree days occur earlier in the year. Furthermore, a blended wintertime shoulder season overlaps with short-lived peak demand events from winter storms. Under this scenario there could be at least 1.75\% or as high as 23.1\% chance of demand exceeding supply.

Our results highlight the effects that climate change might have on electrical grid reliability, and the importance of taking climate change into account in planning for the future. We expect similar shifts in shoulder season timing to occur in other electricity markets around the world.

\section{Acknowledgments}
This work was partially supported by the Texas State Energy Conservation Office, U.S. Army Corps of Engineers ERDC (Energy Research and Development Center), Artesion Inc., Meta, Grid United, and The Lemelson Foundation. ERCOT generation and outage data were obtained from \url{https://gridstatus.io}.

\section{Author contributions}
Conceptualization, H.D.; Methodology, H.D. and J.D.R.; Investigation, H.D. and A.P.; Writing -- Original Draft, H.D. and J.D.R.; Writing -- Review \& Editing, H.D., J.D.R., and M.E.W.; Supervision, H.D.

\section{Declaration of interests}
Webber and Rhodes receive funding from a variety of government agencies, foundations, and industry, including financial institutions and energy companies. A full list of supporters is available at \url{http://www.webberenergygroup.com} and list of public consulting reports, including the funders of those reports, is available at \url{https://www.ideasmiths.net/reports-publications/}. Webber is on the board of GTI Energy, CTO at Energy Impact Partners, and co-founder and Chairman of IdeaSmiths LLC, an engineering consulting firm. Rhodes is on the board of Catalyst Cooperative, and co-founder and CTO of IdeaSmiths LLC. Any opinions, findings, conclusions or recommendations expressed in this material are those of the authors and do not necessarily reflect the views of the sponsors, Energy Impact Partners, GTI Energy, Catalyst Cooperative, The University of Texas at Austin, Columbia University, or IdeaSmiths LLC. The terms of this arrangement have been reviewed and approved by the University of Texas at Austin in accordance with its policy on objectivity in research.

Daigle receives funding from a variety of government agencies, foundations, and industry. Any opinions, findings, conclusions or recommendations expressed in this material are those of the authors and do not necessarily reflect the views of his sponsors.

\appendix

\section{Supporting Information}
\subsection{Calculation of percentage of unmet demand with additional planned outages (Table 2)}

In Table 2, we present the percentage of demand that would not be met in December and January for the years 2020-2022 if an additional 5.5 GW of planned outages for power plant maintenance had occurred. To obtain this value, we subtracted 5.5 GW from the maximum telemetered output for the month in question. We then determined the fraction of hourly peak demand that exceeded this value during that month. Note that this calculation does not attempt to match supply and demand hour for hour; we treat telemetered output and peak demand as random and simply calculate the probability that peak demand during any hour would not be met.

\begin{table}
    \centering
    \begin{tabular}{ |c|c|c|c|c|c| }
        \hline
        Year & $a_1$ & $a_2$ & $a_3$ & $a_4$ & $T_0$ ($^\circ$C) \\
        \hline
        1996 & 0.9044 & 25.70 & -1263 & 36850 & 14.09 \\
        \hline
        1997 & 1.606 & 4.216 & -1169 & 38050 & 14.73 \\
        \hline
        1998 & 0.4149 & 55.48 & -1765 & 40560 & 13.78 \\
        \hline
        1999 & -0.04506 & 87.09 & -2304 & 43540 & 13.37 \\
        \hline
        2000 & -0.2412 & 98.86 & -2552 & 46660 & 13.58 \\
        \hline
        2002 & 1.285 & 29.02 & -1644 & 44080 & 14.45 \\
        \hline
        2003 & 0.7175 & 56.84 & -2026 & 46100 & 14.08 \\
        \hline
        2004 & 2.236 & -9.091 & -1261 & 44810 & 15.13 \\
        \hline
        2005 & 0.6223 & 71.55 & -2469 & 50480 & 14.51 \\
        \hline
        2006 & 0.02227 & 94.8 & -2644 & 50870 & 13.88 \\
        \hline
        2007 & 1.694 & 20.79 & -1784 & 49590 & 15.09 \\
        \hline
        2008 & 0.6815 & 68.94 & -2500 & 53220 & 14.86 \\
        \hline
        2009 & 0.6481 & 70.95 & -2528 & 52610 & 14.81 \\
        \hline
        2010 & 1.36 & 32.19 & -1909 & 50780 & 15.14 \\
        \hline
        2011 & 0.7497 & 54.08 & -2086 & 51640 & 14.76 \\
        \hline
        2012 & 0.1019 & 97.31 & -2810 & 54600 & 14.13 \\
        \hline
        2013 & 0.7758 & 69.16 & -2568 & 55800 & 14.85 \\
        \hline
        2014 & 1.706 & 26.19 & -2017 & 55590 & 15.38 \\
        \hline
        2015 & 0.972 & 66.39 & -2658 & 58630 & 15.05 \\
        \hline
        2016 & 0.0002521 & 118.7 & -3468 & 62700 & 14.6 \\
        \hline
        2017 & 1.151 & 55.11 & -2374 & 58030 & 14.73 \\
        \hline
        2018 & 1.496 & 41.03 & -2342 & 60690 & 15.47 \\
        \hline
        2019 & 0.4101 & 104.2 & -3411 & 66680 & 15.04 \\
        \hline
        2020 & -1.632 & 221.5 & -5508 & 78980 & 14.87 \\
        \hline
        2021 & 3.013 & -33.43 & -1112 & 56520 & 15.39 \\
        \hline
        2022 & 1.018 & 54.93 & -2313 & 65140 & 14.89 \\
        \hline
    \end{tabular}
    \caption{Polynomial coefficients used to determine $T_0$ by year. The polynomial has the form $D = a_1T_{avg}^3 + a_2T_{avg}^2 + a_3T_{avg} + a_4$, where $D$ is daily peak demand in MW and $T_{avg}$ is daily average temperature.}
\end{table}

\begin{figure}
    \centering
    \includegraphics{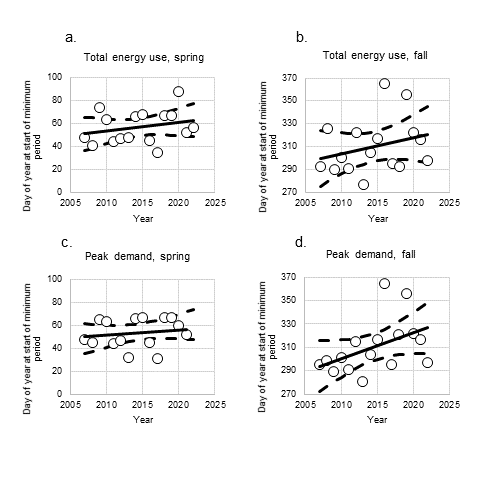}
    \caption{Shoulder seasons defined by electricity with energy and power demand met by non-thermal sources removed. (a) Spring shoulder season defined by total energy use. (b) Spring shoulder season defined by peak demand. (c) Fall shoulder season defined by total energy use. (d) Fall shoulder season defined by peak demand. Linear regression lines with 95\% confidence intervals are shown.}
    \label{figS1}
\end{figure}

\begin{figure}
    \centering
    \includegraphics{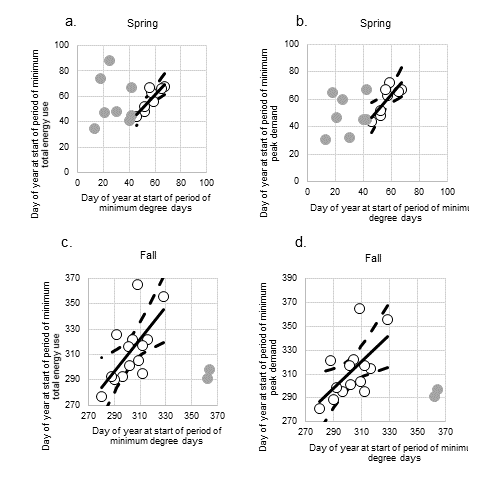}
    \caption{Correlations between start of 45-day periods of minimum degree days, minimum total energy use, and minimum peak demand during spring (a, b) and fall (c, d). Electricity data have energy and power demand met by non-thermal generation removed. The gray data points occur before or after the dates mentioned in the text.}
    \label{figS2}
\end{figure}


\bibliographystyle{elsarticle-harv} 
\bibliography{cas-refs}





\end{document}